\documentclass[epj]{svjour}
\usepackage{graphicx}
\usepackage{color}
\usepackage{amsmath}
\usepackage{amssymb}
\usepackage{jjmacros}
\usepackage{version}
\begin{document}
\title{Photoproduction of \boldmath$\piz\omega$\unboldmath\ off protons
for \boldmath$E_{\gamma}\leq 3$\unboldmath\,GeV}
\author{
The CB-ELSA Collaboration \medskip \\
J.~Junkersfeld$\,^1$,
A.V.~Anisovich$\,^{1,2}$,
G.~Anton$\,^3$,
R.~Bantes$\,^4$,
O.~Bartholomy$\,^1$,
R.~Beck$\,^1$,
Yu.~Beloglazov$\,^2$,
R.~Bogend\"orfer$\,^3$,
R.~Castelijns$\,^{5,\rm a}$,
V.~Crede$\,^{1,6}$,
A.~Ehmanns$\,^1$,
J.~Ernst$\,^1$,
I.~Fabry$\,^1$,
H.~Flemming$\,^{8,\rm b}$,
A.~F\"osel$\,^3$,
M.~Fuchs$\,^1$,
Ch.~Funke$\,^1$,
R.~Gothe$\,^{4,\rm c}$,
A.~Gridnev$\,^{2}$,
E.~Gutz$\,^1$,
St.~H\"offgen$\,^4$,
I.~Horn$\,^1$,
J.~H\"o\ss l$\,^3$,
H.~Kalinowsky$\,^1$,
F.~Klein$\,^4$,
E.~Klempt$\,^1$,
H.~Koch$\,^8$,
M.~Konrad$\,^4$,
B.~Kopf$\,^8$,
B.~Krusche$\,^9$,
J.~Langheinrich$\,^{4,\rm c}$,
H.~L\"ohner$\,^5$,
I.~Lopatin$\,^2$,
J.~Lotz$\,^1$,
H.~Matth\"ay$\,^8$,
D.~Menze$\,^4$,
J.~Messchendorp$\,^{7,\rm d}$,
V.A.~Nikonov$\,^{1,2}$,
D.~Novinski$\,^2$,
M.~Ostrick$\,^{4,\rm e}$,
A.~Radkov$\,^2$,
A.V.~Sarantsev$\,^{1,2}$,
S.~Schadmand$\,^{7,\rm a}$,
C.~Schmidt$\,^1$,
H.~Schmieden$\,^4$,
B.~Schoch$\,^4$,
G.~Suft$\,^3$,
V.~Sumachev$\,^2$,
T.~Szczepanek$\,^1$,
H.~van~Pee$\,^{1,7}$,
U.~Thoma$\,^{1,7}$,
D.~Walther$\,^4$,~and
Ch.~Weinheimer$\,^{1,\rm f}$ }

\institute{$^1\,$Helmholtz-Institut f\"ur Strahlen- und Kernphysik,
Universit\"at Bonn, Germany\\
$^2\,$Petersburg Nuclear Physics Institute, Gatchina, Russia\\
$^3\,$Physikalisches Institut, Universit\"at Erlangen, Germany\\
$^4\,$Physikalisches Institut, Universit\"at Bonn, Germany\\
$^5\,$Kernfysisch Versneller Instituut,
Groningen, The Netherlands\\
$^6\,$Department of Physics, Florida State University, Tallahassee, FL,
USA\\
$^7\,$II. Physikalisches Institut, Universit\"at
Gie{\ss}en, Germany\\
$^8\,$Institut f\"ur Experimentalphysik I, Universit\"at Bochum,
Germany\\
$^9\,$Institut f\"ur Physik, Universit\"at Basel,
Switzerland\\[1mm]
$^{\rm a}\,$Present address: Institut f\"ur Kernphysik,
Forschungszentrum J\"ulich, Germany\\ 
$^{\rm b}\,$Present address: GSI, Darmstadt, Germany\\
$^{\rm c}\,$Present address:
University of South Carolina, Columbia, SC, USA\\ $^{\rm d}\,$Present
address: Kernfysisch Versneller Instituut, Groningen, The Netherlands\\
$^{\rm e}\,$Present address: Institut f\"ur Kernphysik, Universit\"at Mainz, Germany\\
$^{\rm f}\,$Present address: Institut f\"ur Kernphysik, Universit\"at M\"unster, Germany\\
}
\date{Received: \today / Revised version:}

\abstract{
Differential and total cross-sections for photoproduction of
$\rm\gamma p\to p\piz\omega$ and $\rm\gamma p\to\Delta^{+}\omega$
 were determined from measurements of the CB-ELSA experiment, performed at the
 electron accelerator ELSA in Bonn. The measurements covered the photon energy
 range from the production threshold up to $3\,\GeV$.
 % end of PACS codes
} %end of abstract
\authorrunning{J. Junkersfeld {\it et al.}}
\titlerunning{Photoproduction of $\piz\omega$
off protons}

\mail{klempt@hiskp.uni-bonn.de}

\maketitle
\section{Introduction}
\label{intro}
A large number of baryon resonances has been established
experimentally~\cite{PDG}. Below a mass of 1.8\,GeV/c$^2$, most of these states
are well reproduced by constituent quark models
\cite{Capstick:bm,Riska,Metsch}. The models differ in
details of the predicted mass spectrum but have a common feature:
above $1.8\,\GeV/{\rm c}^2$, they predict many more states than have been seen
experimentally. A natural explanation for these \emph{missing
resonances} is that they have escaped detection. The majority of
known non-strange baryon resonances stems from $\pi\text{N}$ scattering
experiments. Model calculations show that for some of these missing
resonances only a small coupling to $\pi\text{N}$ is
expected~\cite{Capstick:1993kb}. In elastic scattering, the coupling to
$\pi\text{N}$  enters in the entrance and exit channel so these
resonances contribute only very weakly.
By contrast, these resonances are predicted to have normal photo couplings
\cite{Capstick:1992uc} and some of them should be observed in
channels like $\rm N\eta$, $\rm K\Lambda$, $\rm K\Sigma$
\cite{Capstick}, $\rm\Delta\eta$ or $\rm\Delta\omega$
\cite{Capstick:1998md}. In comparison to the $\rm N\pi$ final state,
most of the above provide a distinctive advantage: they act as isospin
filters; only $\text{N}^{*}$ resonances contribute to the $\rm N\eta$
and $\rm K\Lambda$ final states while resonances in $\rm\Delta\eta$,
and $\rm\Delta\omega$   belong to the $\rm \Delta^*$
series.

A partial-wave analysis of various photoproduction data
suggested the existence of several new resonances~\cite{sarantsev}. The
analysis included data from CB-ELSA on $\pi^0$ and $\eta$
photoproduction \cite{Bartholomy:04,Crede:04}, Mainz-TAPS data
 on $\eta$ photoproduction~\cite{Krusche:nv}, beam-asym\-metry
measurements of $\pi^0$ and $\eta$~\cite{GRAAL1,SAID1,GRAAL2}, data on
$\rm\gamma p\rightarrow n\pi^+$~\cite{SAID2} and from the compilation
of the SAID database \cite{SAID}, and data on photoproduction of
$\rm\gamma p\to K^+\Lambda$ and $\rm\gamma p\to K^+\Sigma^0$ from
SAPHIR~\cite{Glander:2003jw,Lawall:2005np},
CLAS~\cite{McNabb:2003nf,carnahan}, and LEPS~\cite{Zegers:2003ux}.

\begin{figure}[t]
\begin{tabular}{cc}
\includegraphics[width=0.23\textwidth]{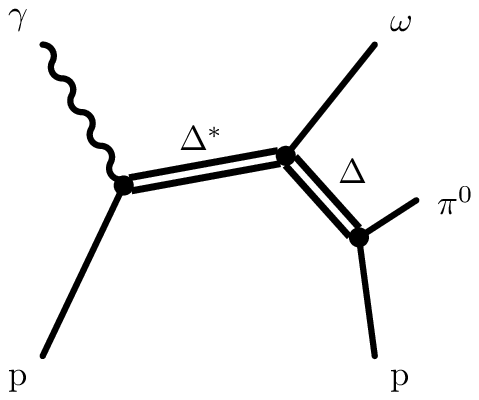}&
\includegraphics[width=0.23\textwidth]{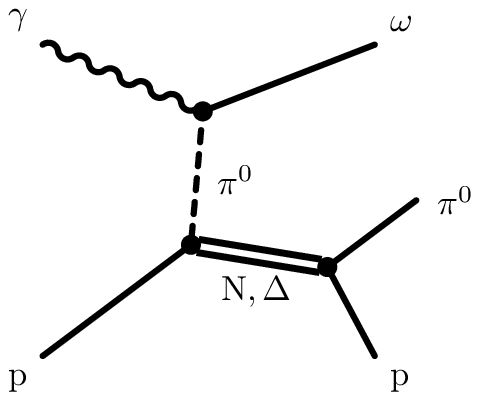}
\end{tabular}
\caption{\label{feynm}
Contributions to $\rm\Delta\omega$ photoproduction: left, production of
$\rm \Delta^*$ intermediate states; right:  production of $\omega$ mesons via $t$-channel pion exchange.
} \end{figure}

The reaction $\rm \gamma p\to p\omega$ is known to receive large 
contributions from $t$-channel exchange processes~\cite{Barth:om,Ajaka:2006bn}. 
A similar mechanism may contribute also to $\rm\Delta\omega$
photoproduction: the incoming photon may couple to $\omega\pi^0$, the
virtual $\pi^0$ excites the nucleon to a $\rm\Delta$ and the $\omega$
escapes, preferentially in forward direction. Fig.~\ref{feynm} shows
a Feynman diagram for this reaction mechanism and for the production of a $\rm\Delta$ resonance decaying into $\rm\Delta\omega$.

This paper reports on a measurement of differential and total cross
sections for the reaction
\begin{equation}
\label{eqn:reac.ppi0omega}
\gpppizomega \, , 
\end{equation}
with $\omega\to\piz\gamma$ and the $\pi^{0}$ detected in its two photon decay. From this data
the total cross-section for
\begin{equation}
\gpdeltaomega%\qquad\rm\Delta^{+}\to\proton\pi^0%\qquad
%\omega\to\pi^{0}\gamma\, 
\label{eqn:reac.deltaomega} \end{equation}
with the subsequent decays \[\rm\Delta^{+}\to\proton\pi^0\qquad \text{and}\qquad  \omega\to\pi^{0}\gamma\]
was extracted and compared to an earlier measurement at higher
energies~\cite{Barber:1984fr}.
The low statistics for reactions~(\ref{eqn:reac.ppi0omega})
and~(\ref{eqn:reac.deltaomega})  does not yet provide a sufficiently
large data sample for a partial-wave analysis, but  may serve as a
guide for what to expect from future experiments and is thus of exploratory
character.

\section{Experimental setup}
\label{sec:setup}

\begin{figure}[t]
\begin{center}
\includegraphics[width=0.48\textwidth]{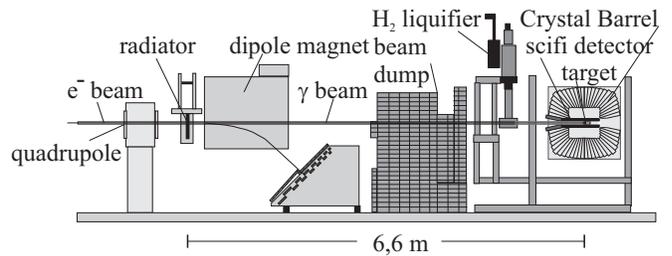} \end{center}
\caption{\label{fig:exp.setup}
Setup of the CB-ELSA experiment}
\vspace{-4mm}
\end{figure}

The experiment was performed at the Electron Stretcher Accelerator ELSA~\cite{Hillert:2006yb}
at the University of Bonn. Electrons were extracted at an energy of
$3.2\,\GeV$ and brems\-strahlung was produced in a radiator foil with a
thickness of $3/1000$ of a radiation length (Fig.~\ref{fig:exp.setup}).
Electrons deflected  in a magnet were detected with a tagging system
covering the photon energy range from $750$ to $2970\,\MeV$. The
tagging system  consisted of 14 thick scintillation counters and two
proportional wire chambers with a total of 352 wires. The scintillation
counters were used to derive a fast timing signal and the wire chambers
to determine the photon energy.
The $\gamma$ energy resolution varied from  30\,MeV at the lower end to 0.5\,MeV at the
upper end of the spectrum not taking 
into account the energy distribution of the
electron beam of $~3-5\,\MeV$~\cite{Hillert:2006yb}. This is  well matched
with the overall resolution of the detector for this
reaction of $30\,\MeV$ (FWHM) (see below).
Typical rates were $1-3 \times 10^6 \,
\text{photons}/\textrm{s}$. The photon beam hit a liquid $\rm{H}_2$
target of $5.3\,\rm{cm}$  length and $3\,\rm{cm}$  diameter. The
absolute normalisation was derived from a comparison of our
differential angular distributions for the reaction $\rm\gamma p\to
p\pi^0$ with the SAID model SM02. The normalisation uncertainty was
estimated to be $15\,\%$ \cite{Bartholomy:04,Pee:06}.

Charged reaction products were detected by a three-layer scintillating
fibre  detector covering polar angles from $15\degree$ to $165\degree$
\cite{Suft:2005cq}. The outer layer was parallel to the beam axis, the fibres of the
other two layers were bent $\pm 25\degree$ with respect to the first
layer  to allow for a spatial reconstruction of hits. Photons and
charged particles were detected in the Crystal Barrel detector
\cite{Aker:1992ny}, a calorimeter consisting of 1380 $\rm{CsI(Tl)}$
crystals with photodiode readout, covering $98\,\%$ of $4\pi$ solid
angle. The detector with its high granularity and energy resolution  is
excellently suited for the detection of multi-photon final states.

Electromagnetic showers typically extended  over up to 30 crystals in
the calorimeter. Photons were reconstructed with an energy resolution
of $\sigma_E/E = 2.5\% / \sqrt[4]{E [\GeV]}$ and an angular
resolution of $\sigma_{\theta,\phi} \approx 1.1^{\circ}$. Hits due to
charged particles induce smaller clusters with typical 3 -- 6
crystals. 

A fast first-level trigger signal was derived from a coincidence
between a hit in the tagging  system and a signal in at least two out
of three layers of the inner fibre detector. The second-level trigger
required a minimum number of hits in the calorimeter. For part of the
data the minimal number of hits in the calorimeter was 2, otherwise at least 3
hits were requested, in order to reduce the dead-time. Dead-time losses were
below $70\,\%$, and below $20\,\%$ for the more restrictive trigger.

A more detailed description of the experimental setup and  the
event reconstruction 
can be found
in~\cite{Pee:06}. 

\section{The reaction \boldmath$\gp\to\ppizomega$\unboldmath}
\subsection{Event selection}

The reaction $\gp\to\proton\piz\omega$, $\omega\to\piz\gamma$, leads to
a final state with five photons and a proton. The $\piz\omega$
photoproduction threshold is at $E_{\gamma} = 1365$\,MeV; a cut
on a tagged photon energy $E_{\gamma}>1315$\,MeV
was applied right at the beginning.

The first step in the analysis is the identification of the five
photons and the reconstruction of their energies and directions. 
Protons  with $E_{\text{kin}}$ below
$\sim 95\,\MeV$ only produce a signal in the inner detector but not in
the calorimeter. Hence in the analysis, events were selected with 5 or
6 hits in the Crystal Barrel calorimeter and 1 -- 3 hits in the inner
detector. (A three--hit pattern can arise from three single hits in
each layer not crossing in a single point.) At least two layers of the
inner detector had to have a signal. For each pair of fibre and barrel
hits it was tested if the two vectors pointing from the target centre to a
fibre--detector-hit and to a Crystal-Barrel-hit form an
angle of $20^{\circ}$ or less; in this case the Crystal Barrel hit was
identified as a proton, otherwise as a photon. The $20^{\circ}$ {\it
matching} angle was chosen to allow for the extension of the target
and the uncertainties in the measurement. Events with five photons were
kept for further analysis. In case of 6 hits in the barrel, one of them
had to match the proton identification.

Surviving events were kinematically fitted to the hypothesis $\rm\gamma
p\to \rm{p}_{\text{miss}} \,\pi^0\pi^0\gamma$ with a missing proton,
neglecting identified charged hits and using all remaining photon
candidates. The kinematic fit assumed that the reaction took place in
the target centre. Since the momentum of the proton is unknown and
needs to be reconstructed, energy and momentum conservation give one
constraint, the $\pi^0$ masses two constraints. A cut on a confidence
level of $2\,\%$ was applied, optimised to lose only few good events. From the fit, the flight direction of
the proton was determined and compared to hits in the inner
detector. Again, the direction of the missing proton and the
direction to a hit in the inner fibre detector had to
form an angle of $\pm 20^{\circ}$ or less for the hit 
to be  identified
as proton.

\begin{figure}[t]
\begin{center}
\includegraphics[width=0.4\textwidth]{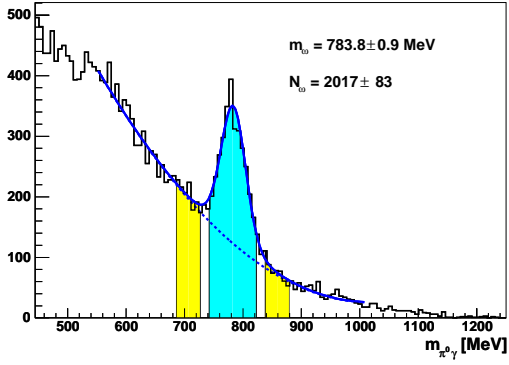}
\end{center}
\caption{\label{fig:mpi0g.sig}
$\omega$ signal in $\piz\gamma$ invariant mass}
\end{figure}

The $\proton\piz\piz\gamma$ events were used to identify
$\proton\piz\omega$ events with $\omega$ decaying into $\piz\gamma$.
Fig.~\ref{fig:mpi0g.sig} shows the $\pi^0\gamma$ mass distribution with
two entries per event. The fit using a Voigt function (a Breit-Wigner
convoluted with a Gaussian) imposing the $\omega$ width of
$\Gamma=8.49$\,MeV/c$^2$  assigns about 2000 events to
reaction~(\ref{eqn:reac.ppi0omega}). The $\omega$ mass was determined to $(783.8\pm
0.9_{\text{stat}}\pm 1.0_{\text{syst}})$\,MeV/c$^2$. 
The systematic error
was estimated from the comparison of $\eta, \eta^{\prime}$, and $\omega$
masses in different reactions with the
PDG values. The mass
resolution is determined to $\sigma=16$\,MeV/c$^2$.

\begin{figure}[t]
\begin{tabular}{cc}
\hspace{-5mm}\includegraphics[width=0.25\textwidth]{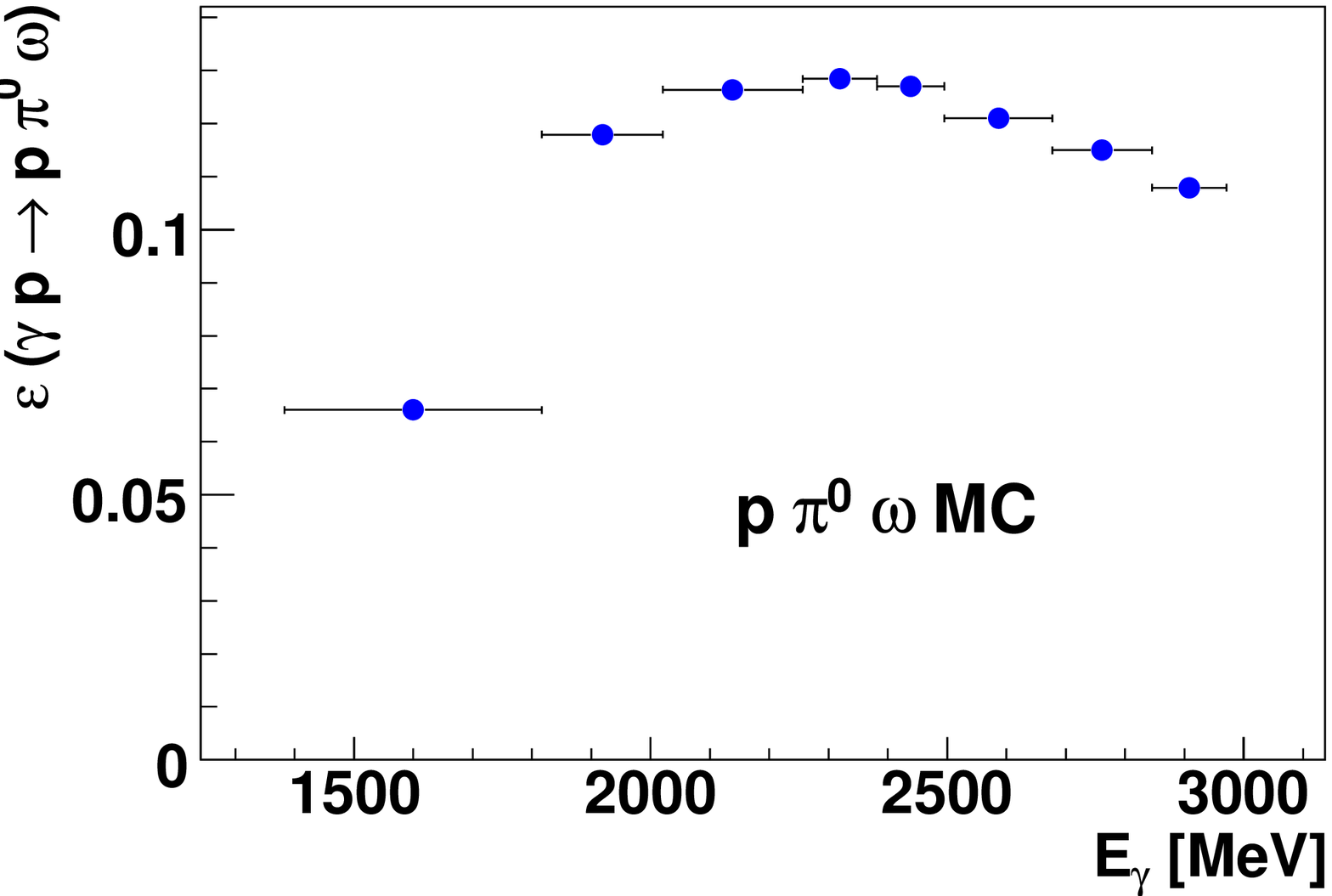}&
\hspace{-4mm}\includegraphics[width=0.25\textwidth]{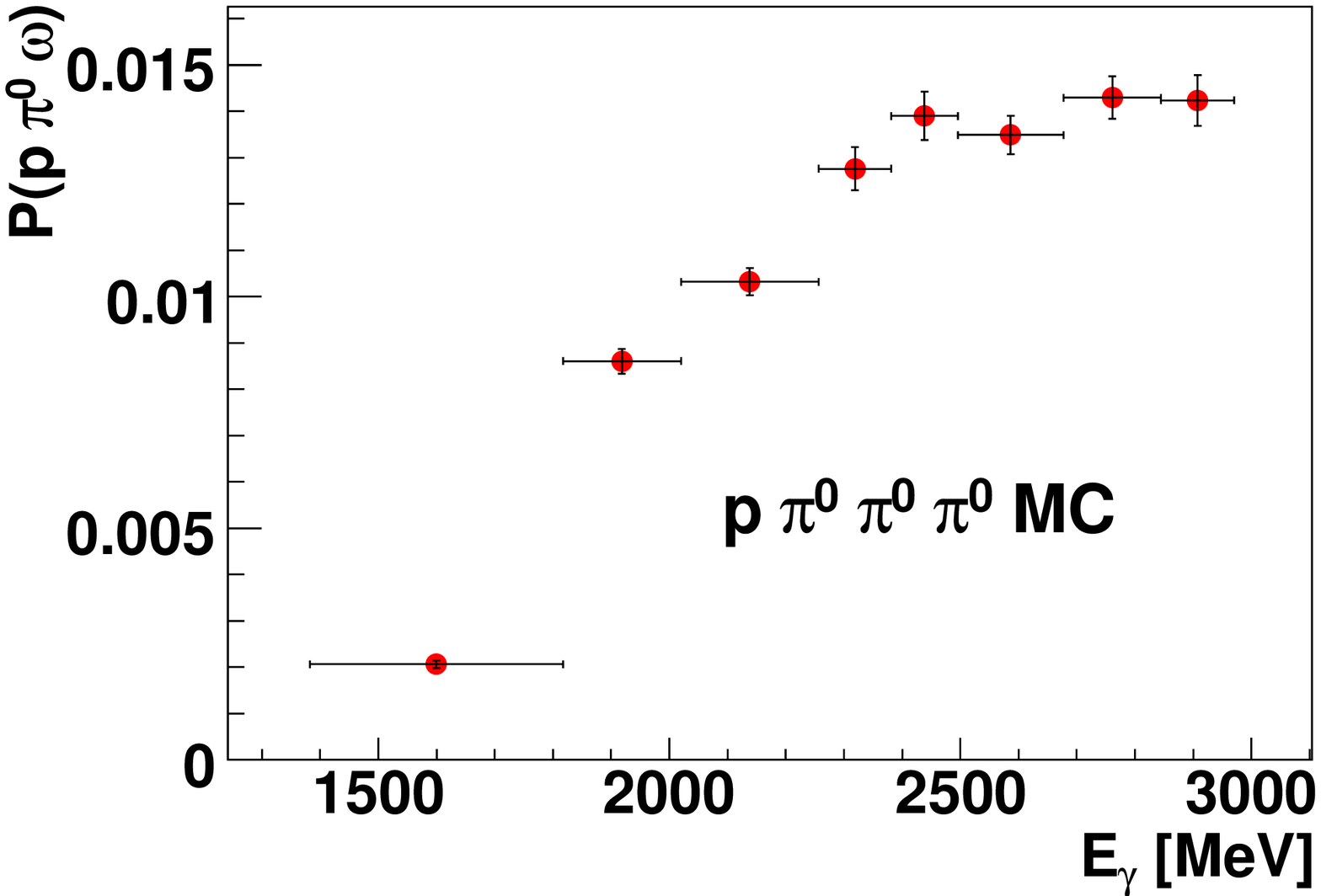}
\end{tabular}
\caption{\label{fig:acc}
Acceptance of $\proton\piz\omega$ events (left) and the
misidentification probability of $\proton\, 3\piz$ events (right).}
%\end{figure}
%\begin{figure}
\begin{tabular}{cc}
\hspace{-3mm}\includegraphics[width=0.25\textwidth]{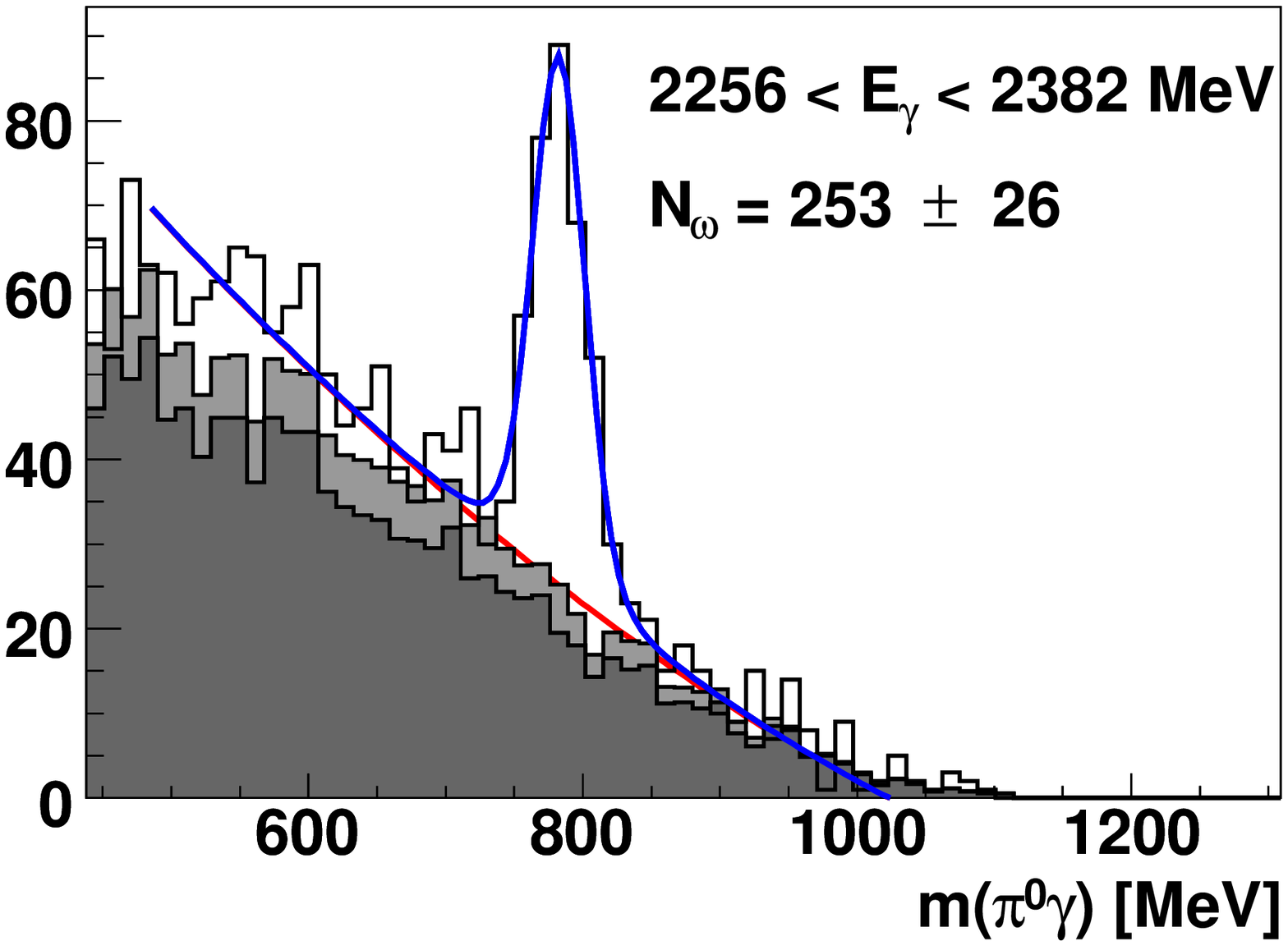}&
\hspace{-6mm}\includegraphics[width=0.25\textwidth]{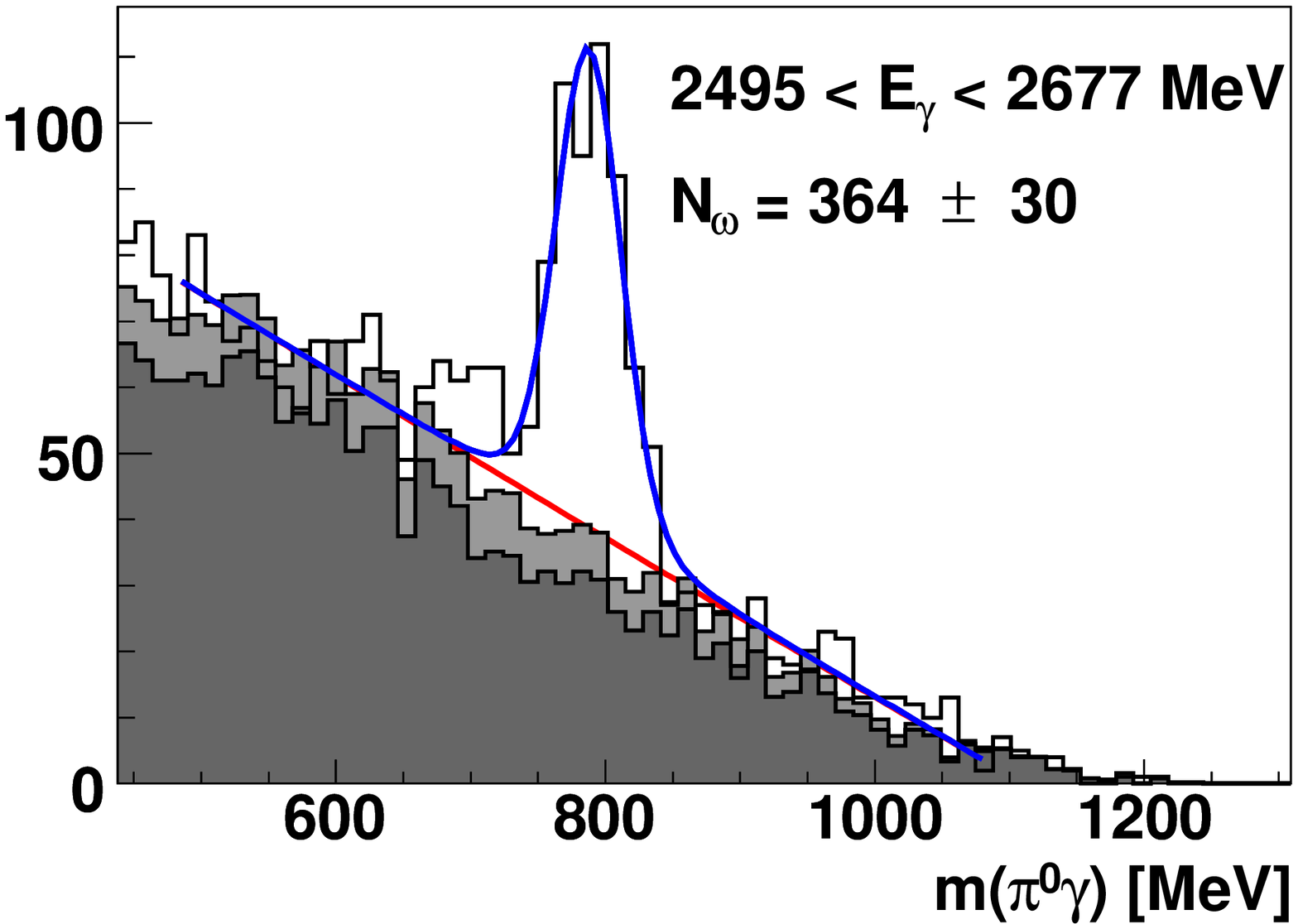}
\end{tabular}
\caption{\label{fig:mpi0g.bg}
The $\omega$ signal with background. The background is predicted in height and shape by simulations of  $\proton\, 3\piz$ (dark-grey) 
and $\ppizomega$ combinatorial background (light-grey).}
\end{figure}

The background in the $\piz\gamma$ distribution of
$\proton\piz\piz\gamma$ events has two main sources. A large fraction
stems from $\proton\, 3\piz$ events. Fig.~\ref{fig:acc} shows that, in
the energy region $2000 < E_{\gamma} < 3000\,\MeV$, $\proton\, 3\piz$
events have a high probability to be misidentified as
$\proton\piz\omega$. The misidentification probability is only one
order of magnitude smaller than the acceptance for $\ppizomega$.
However, the branching ratio of $\proton\, 3\piz \to \proton\, 6\gamma$
is $(96.44\pm 0.09)\,\%$ compared with $(8.71 \pm 0.25)\,\%$ for
$\ppizomega\to\proton\, 5\gamma$. The cross-section for $3\pi^0$
photoproduction was estimated using the cross-section for
$\gp\to\proton\eta$~\cite{Crede:04}, which was determined from events
with the $\eta$ decaying into $\gamma\gamma$ and $3\,\pi^{0}$. The
fractions of $\eta$ and non-$\eta$ events in the $3\piz$ event samples
were determined and used to estimate the cross-section of
$\gp\to\proton \, 3\pi^{0}$. Monte Carlo simulations were performed
using the $\proton\, 3\pi^0$ cross-section estimate to determine the
expected number of $\proton\, 3\pi^0$ events surviving the
$\proton\piz\piz\gamma$ reconstruction. Fig.~\ref{fig:mpi0g.bg} shows
for two photon energy ranges the predicted contribution of
$\proton\piz\piz\piz$ events to the background and the observed
$\piz\gamma$ distribution.

The expected combinatorial background was determined from the number of
reconstructed $\proton\piz\omega$ events. It is shown together with
the $\proton\, 3\piz$ part of the background in
Fig.~\ref{fig:mpi0g.bg}. In the $\omega$ mass region, there is good
agreement between the simulated background distribution and the
observed background. The study of simulated $\proton\piz\piz$ and
$\proton\piz\eta$ events shows misidentification probabilities of the
order of $0.1\,\%$. Their contributions were neglected.

The number of events due to reaction~(\ref{eqn:reac.ppi0omega}) in a
given energy range  was determined by fitting
the $\pi^0\gamma$ distribution using  a Voigt function for the $\omega$
signal and a second order polynomial for the background. The fit also
returned the number of background events below the peak. For background
subtraction, data histograms were filled with events within the
$\omega$ mass region ($m_{\omega}\pm 40\,\MeV/{\text{c}^2}$) and
background histograms with events falling into the upper or lower
sidebands  ($687 - 727\,\MeV/{\text{c}^2}$ and $839 -
879\,\MeV/{\text{c}^2}$, also shown in Fig.~\ref{fig:mpi0g.sig}). The latter histograms were scaled to contain
the same number of events as found in the background below the peak.
For each energy and angular region, the sideband histograms were
subtracted from the data histograms to extract the $\ppizomega$
distributions. The same procedure was used to determine the $\ppizomega$
distributions for each energy region as function of the momentum transfer and the invariant mass respectively.

The acceptance was studied with a GEANT-based Mon\-te Carlo simulation
using  phase space distributed $\ppizomega$ and $\rm\Delta^{+}\omega$
events. In the first iteration only $\ppizomega$ Monte-Carlo events
were taken into account. cross-sections for $\gpppizomega$ and
$\gpdeltaomega$ were thus obtained, as will be described in sections~\ref{sec:cs.total} and~\ref{sec:deltaomega}, and used to produce Monte Carlo events
with a realistic mixture of $\proton\piz\omega$ and
$\rm\Delta^{+}\omega$ events. This provides a more realistic acceptance simulation. 
Stable results were achieved in the
second iteration.
The simulated acceptance was different when only 
events due to phase-space distributed $\ppizomega$ events or $\rm\Delta^{+}\omega$ events  were used for the simulation. 
The difference in the acceptance %of less than $5\,\%$ 
was taken as a contribution to the systematic error.

\subsection{Differential cross-sections}
\label{sec:cs.diff}

\begin{table*}[tp]
\caption{\label{tab:cs.cosThOmega}
Differential cross-sections $\dsdomegatext (\cos\theta_{\omega})$. There is a common systematic error of $16\,\%$.}
\begin{center}
\begin{tabular}{ccccc}
\hline
$\cos \theta_{\omega}$      & $\dsdomegatext (\cos\theta_{\omega})$&  $\dsdomegatext (\cos\theta_{\omega})$& $\dsdomegatext
(\cos\theta_{\omega})$&   $\dsdomegatext (\cos\theta_{\omega})$ \\
                           & $[\ub/\text{sr}]$ & $[\ub/\text{sr}]$ & $[\ub/\text{sr}]$ & $[\ub/\text{sr}]$ \\
\hline
$E_{\gamma}$ $[\MeV]$      & 1383 - 1817&           1817 - 2020&         2020 - 2256        & 2256 - 2382\\       
\hline                                                                                                     
$ -1.00 - -0.80$ & $ 0.16 \pm 0.08$ & $ 0.00 \pm 0.13$ & $ 0.29 \pm 0.14$ & $ 0.35 \pm 0.23$ \\ 
$ -0.80 - -0.60$ & $ 0.08 \pm 0.06$ & $ 0.05 \pm 0.07$ & $ 0.18 \pm 0.09$ & $ 0.23 \pm 0.13$ \\ 
$ -0.60 - -0.40$ & $ 0.04 \pm 0.04$ & $ 0.12 \pm 0.06$ & $ 0.15 \pm 0.06$ & $ 0.33 \pm 0.12$ \\ 
$ -0.40 - -0.20$ & $ 0.02 \pm 0.03$ & $ 0.03 \pm 0.04$ & $ 0.05 \pm 0.04$ & $ 0.22 \pm 0.08$ \\ 
$ -0.20 - 0.00$  & $ 0.03 \pm 0.02$ & $ 0.17 \pm 0.04$ & $ 0.16 \pm 0.05$ & $ 0.10 \pm 0.08$ \\ 
$ 0.00 - 0.20$   & $ -0.02 \pm 0.02$ & $ 0.09 \pm 0.05$ & $ 0.17 \pm 0.05$ & $ 0.31 \pm 0.09$ \\ 
$ 0.20 - 0.40$   & $ 0.06 \pm 0.03$ & $ 0.13 \pm 0.04$ & $ 0.22 \pm 0.06$ & $ 0.30 \pm 0.08$ \\ 
$ 0.40 - 0.60$   & $ 0.03 \pm 0.03$ & $ 0.19 \pm 0.06$ & $ 0.33 \pm 0.06$ & $ 0.39 \pm 0.11$ \\ 
$ 0.60 - 0.80$   & $ 0.04 \pm 0.03$ & $ 0.32 \pm 0.06$ & $ 0.46 \pm 0.08$ & $ 0.71 \pm 0.13$ \\ 
$ 0.80 - 1.00$   & $ 0.11 \pm 0.04$ & $ 0.32 \pm 0.07$ & $ 0.75 \pm 0.10$ & $ 0.65 \pm 0.17$ \\ 
\hline
$E_{\gamma}$ $[\MeV]$      & 2382 - 2495          & 2495 - 2677        & 2677 - 2845        & 2845 - 2970\\
\hline                                                                                                      
$ -1.00 - -0.80$ & $ -0.10 \pm 0.25$ & $ 0.27 \pm 0.18$ & $ 0.70 \pm 0.28$ & $ 0.47 \pm 0.23$ \\ 
$ -0.80 - -0.60$ & $ 0.27 \pm 0.14$ & $ 0.44 \pm 0.13$ & $ 0.37 \pm 0.16$ & $ 0.42 \pm 0.16$ \\ 
$ -0.60 - -0.40$ & $ 0.44 \pm 0.13$ & $ 0.27 \pm 0.09$ & $ 0.15 \pm 0.10$ & $ 0.13 \pm 0.12$ \\ 
$ -0.40 - -0.20$ & $ 0.21 \pm 0.10$ & $ 0.17 \pm 0.08$ & $ 0.30 \pm 0.09$ & $ 0.25 \pm 0.11$ \\ 
 $ -0.20 - 0.00$ & $ 0.17 \pm 0.09$ & $ 0.09 \pm 0.07$ & $ 0.14 \pm 0.09$ & $ 0.25 \pm 0.10$ \\ 
  $ 0.00 - 0.20$ & $ 0.25 \pm 0.09$ & $ 0.28 \pm 0.08$ & $ 0.22 \pm 0.08$ & $ 0.17 \pm 0.11$ \\ 
  $ 0.20 - 0.40$ & $ 0.48 \pm 0.11$ & $ 0.26 \pm 0.08$ & $ 0.42 \pm 0.10$ & $ 0.32 \pm 0.12$ \\ 
  $ 0.40 - 0.60$ & $ 0.45 \pm 0.13$ & $ 0.42 \pm 0.10$ & $ 0.42 \pm 0.11$ & $ 0.51 \pm 0.15$ \\ 
  $ 0.60 - 0.80$ & $ 1.02 \pm 0.18$ & $ 0.57 \pm 0.13$ & $ 0.77 \pm 0.18$ & $ 0.99 \pm 0.24$ \\ 
  $ 0.80 - 1.00$ & $ 1.76 \pm 0.27$ & $ 1.23 \pm 0.20$ & $ 1.86 \pm 0.28$ & $ 1.79 \pm 0.33$ \\ 
\hline
\end{tabular}
\end{center}
\end{table*}

\begin{table*}[tp]
\caption{\label{tab:cs.cosThPi0}
Differential cross-sections $\dsdomegatext (\cos\theta_{\piz})$. There is a common systematic error of $16\,\%$.}
\begin{center}
\begin{tabular}{ccccc}
\hline
$\cos \theta_{\piz}$      & $\dsdomegatext (\cos\theta_{\piz})$&  $\dsdomegatext (\cos\theta_{\piz})$& $\dsdomegatext
(\cos\theta_{\piz})$&   $\dsdomegatext (\cos\theta_{\piz})$ \\
                           & $[\ub/\text{sr}]$ & $[\ub/\text{sr}]$ & $[\ub/\text{sr}]$ & $[\ub/\text{sr}]$ \\
\hline
$E_{\gamma}$ $[\MeV]$    & 1383 - 1817           & 1817 - 2020         & 2020 - 2256         & 2256 - 2382  \\
\hline                                                                                                     
$ -1.00 - -0.80$ & $ 0.05 \pm 0.03$  & $ 0.23 \pm 0.07$  & $ 0.30 \pm 0.07$  & $ 0.40 \pm 0.11$ \\ 
$ -0.80 - -0.60$ & $ 0.05 \pm 0.03$  & $ 0.21 \pm 0.05$  & $ 0.32 \pm 0.06$  & $ 0.27 \pm 0.09$ \\ 
$ -0.60 - -0.40$ & $ 0.07 \pm 0.03$  & $ 0.09 \pm 0.05$  & $ 0.29 \pm 0.06$  & $ 0.40 \pm 0.11$ \\ 
$ -0.40 - -0.20$ & $ 0.04 \pm 0.03$  & $ 0.08 \pm 0.05$  & $ 0.27 \pm 0.07$  & $ 0.30 \pm 0.11$ \\ 
 $ -0.20 - 0.00$ & $ -0.04 \pm 0.03$  & $ 0.17 \pm 0.05$  & $ 0.23 \pm 0.06$  & $ 0.42 \pm 0.11$ \\ 
  $ 0.00 - 0.20$ & $ 0.06 \pm 0.03$  & $ 0.17 \pm 0.05$  & $ 0.18 \pm 0.06$  & $ 0.17 \pm 0.10$ \\ 
  $ 0.20 - 0.40$ & $ 0.03 \pm 0.03$  & $ 0.11 \pm 0.05$  & $ 0.30 \pm 0.06$  & $ 0.40 \pm 0.12$ \\ 
  $ 0.40 - 0.60$ & $ 0.03 \pm 0.03$  & $ 0.16 \pm 0.06$  & $ 0.27 \pm 0.07$  & $ 0.32 \pm 0.10$ \\ 
  $ 0.60 - 0.80$ & $ 0.07 \pm 0.04$  & $ 0.18 \pm 0.06$  & $ 0.24 \pm 0.07$  & $ 0.33 \pm 0.11$ \\ 
  $ 0.80 - 1.00$ & $ 0.04 \pm 0.04$  & $ 0.18 \pm 0.08$  & $ 0.18 \pm 0.10$  & $ 0.64 \pm 0.17$ \\ 
\hline
 $E_{\gamma}$ $[\MeV]$    & 2382 - 2495           & 2495 - 2677        & 2677 - 2845         & 2845 - 2970  \\   
\hline                                                                                                    
$ -1.00 - -0.80$ & $ 0.59 \pm 0.14$  & $ 0.67 \pm 0.11$  & $ 0.67 \pm 0.14$  & $ 0.61 \pm 0.17$ \\ 
$ -0.80 - -0.60$ & $ 0.54 \pm 0.12$  & $ 0.29 \pm 0.10$  & $ 0.37 \pm 0.10$  & $ 0.50 \pm 0.14$ \\ 
$ -0.60 - -0.40$ & $ 0.57 \pm 0.13$  & $ 0.37 \pm 0.10$  & $ 0.26 \pm 0.11$  & $ 0.32 \pm 0.12$ \\ 
$ -0.40 - -0.20$ & $ 0.47 \pm 0.13$  & $ 0.31 \pm 0.10$  & $ 0.34 \pm 0.12$  & $ 0.23 \pm 0.14$ \\ 
 $ -0.20 - 0.00$ & $ 0.58 \pm 0.13$  & $ 0.46 \pm 0.10$  & $ 0.64 \pm 0.12$  & $ 0.43 \pm 0.13$ \\ 
  $ 0.00 - 0.20$ & $ 0.22 \pm 0.12$  & $ 0.12 \pm 0.09$  & $ 0.42 \pm 0.12$  & $ 0.49 \pm 0.13$ \\ 
  $ 0.20 - 0.40$ & $ 0.65 \pm 0.14$  & $ 0.39 \pm 0.10$  & $ 0.61 \pm 0.12$  & $ 0.60 \pm 0.14$ \\ 
  $ 0.40 - 0.60$ & $ 0.28 \pm 0.12$  & $ 0.29 \pm 0.10$  & $ 0.37 \pm 0.13$  & $ 0.58 \pm 0.15$ \\ 
  $ 0.60 - 0.80$ & $ 0.39 \pm 0.15$  & $ 0.39 \pm 0.11$  & $ 0.24 \pm 0.12$  & $ 0.09 \pm 0.16$ \\ 
  $ 0.80 - 1.00$ & $ 0.56 \pm 0.27$  & $ 0.15 \pm 0.21$  & $ 0.26 \pm 0.16$  & $ 0.35 \pm 0.26$ \\ 
\hline
\end{tabular}
\end{center}
\end{table*}

The differential cross-sections were obtained from the sideband
subtracted histograms. We give in the centre of mass system cross-sections differential in $\cos\theta_{\omega}$, 
$\cos  \theta_{\piz}$ and $| t - t_{\min}|$, 
$$
\dsdomegatext (\cos\theta_{\omega}), \quad\
\dsdomegatext (\cos  \theta_{\piz}), \quad\
\dsdttext (| t - t_{\min}|), 
$$
respectively.
Here $t$ is the squared four-momentum transfer from the photon beam to the $\proton\piz$ system given by 
\begin{equation}
 t = q^2 = \left(p_{\gamma} - p_{\omega}\right)^2 
\label{eqn:t.channel}
\end{equation}
and $t_{\min}$ is the minimal momentum transfer imposed by kinematics.

\begin{figure}[t]
\includegraphics[width=0.5\textwidth]{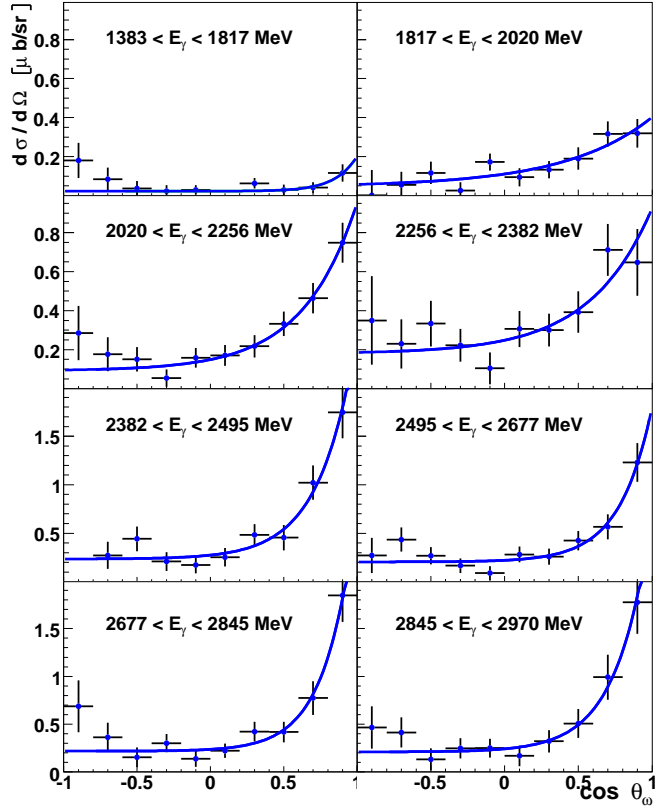}
\caption{\label{fig:cs.cosThOmega}
Differential cross-sections $\dsdomegatext (\cos\theta_{\omega})$.}
\end{figure}

\begin{figure}[b]
\includegraphics[width=0.5\textwidth]{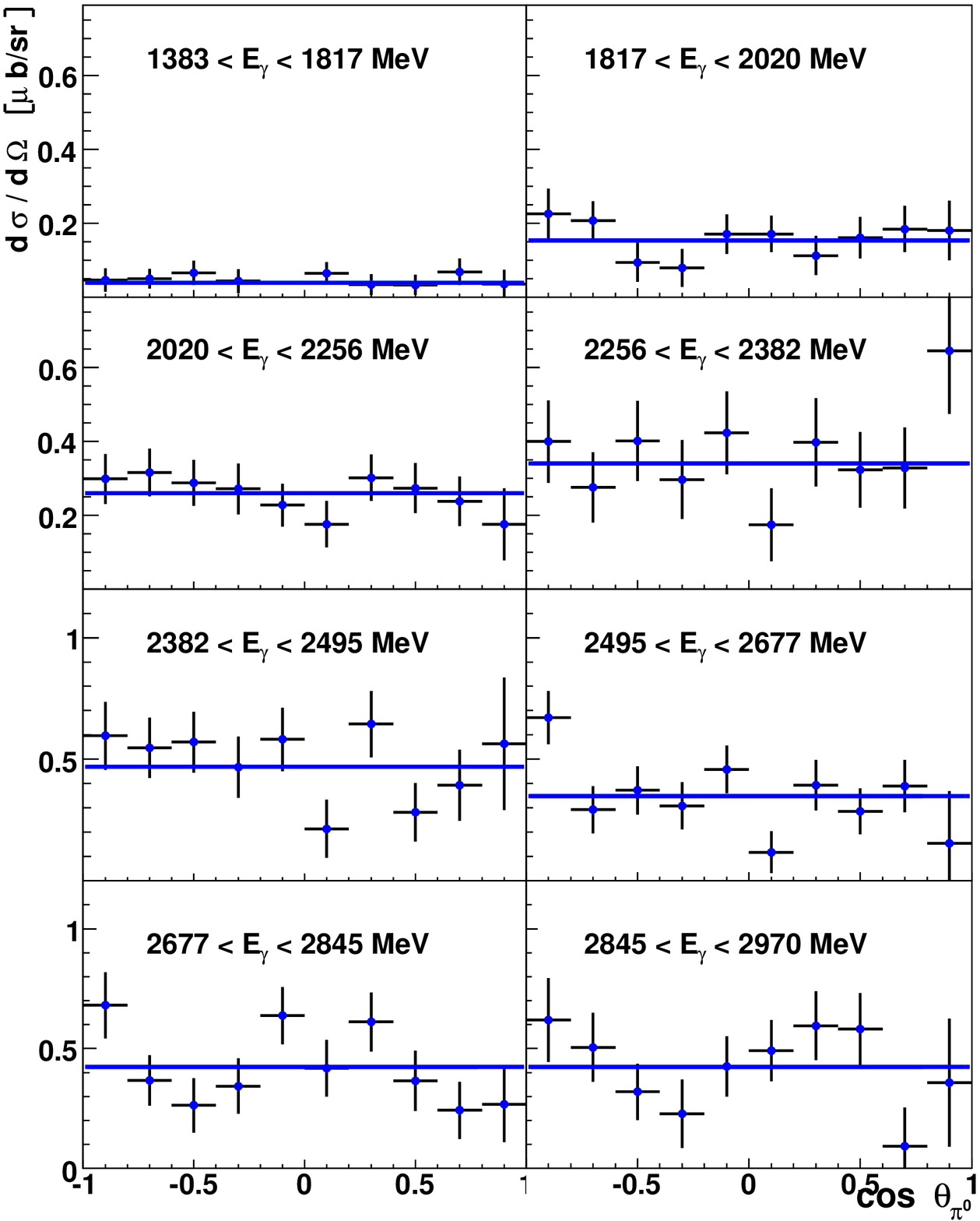}
\caption{\label{fig:cs.cosThPi0}
Differential cross-sections $\dsdomegatext (\cos\theta_{\piz})$. }
\end{figure}

Fig.~\ref{fig:cs.cosThOmega} presents the  differential cross-sections as a
function of $\cos\theta_{\omega}$, in table~\label{tab:cs.cosThOmega} they are given in numerical form. The distributions are compatible
with a description of the form
\begin{equation}
 \dsdomega (x) = a_0 + a_1 \cdot e^{a_2 x} \text{ with } x = \cos \theta.
\label{eqn:fit.dsdomega}
\end{equation}
In the backward direction, the acceptance is small and the errors
large. The fit using  Eq.\ (\ref{eqn:fit.dsdomega}) took into account
only data  for which the acceptance $\epsilon$ was above $5\,\%$ (thus
restricting the fit range to $\cos \theta_{\omega} > -0.6$ for
$E_{\gamma} < 1800\,\MeV$, and to $\cos \theta_{\omega} > -0.8$ for
$E_{\gamma} > 1800\,\MeV$), and was then extrapolated to cover the full
$\cos \theta_{\omega}$ range. In the forward direction, there is a
strong increase in intensity, in particular at energies above
$2\,\GeV$. Production of $\omega$ mesons via $t$-channel exchange with
simultaneous $\rm p\ \to\ \Delta(1232)$ excitation seems to play an
important role in the dynamics of reaction~(\ref{eqn:reac.ppi0omega}).

From the $\cos \theta_{\omega}$ distributions a 
total cross-section was determined by summing over the measured
values for which the acceptance was above $5\,\%$ and using
extrapolated values in the remaining range. 

The differential cross-sections $\dsdomegatext (\cos  \theta_{\piz})$
are shown in Fig.~\ref{fig:cs.cosThPi0} and listed numerically in~\ref{tab:cs.cosThPi0}. There are no obvious
structures beside some fluctuations in the forward
and backward regions. The data were fitted using a constant. The fit was
restricted to data points measured with an acceptance of at least $5\,\%$, thus excluding for $E_{\gamma} > 2380\,\MeV$ the points with $\cos
\theta_{\piz} > 0.8$. From this distribution  
the total cross-section is derived from  the data points 
and the extrapolation was used for the points with small acceptance.

\begin{figure}[tb]
\includegraphics[width=0.5\textwidth]{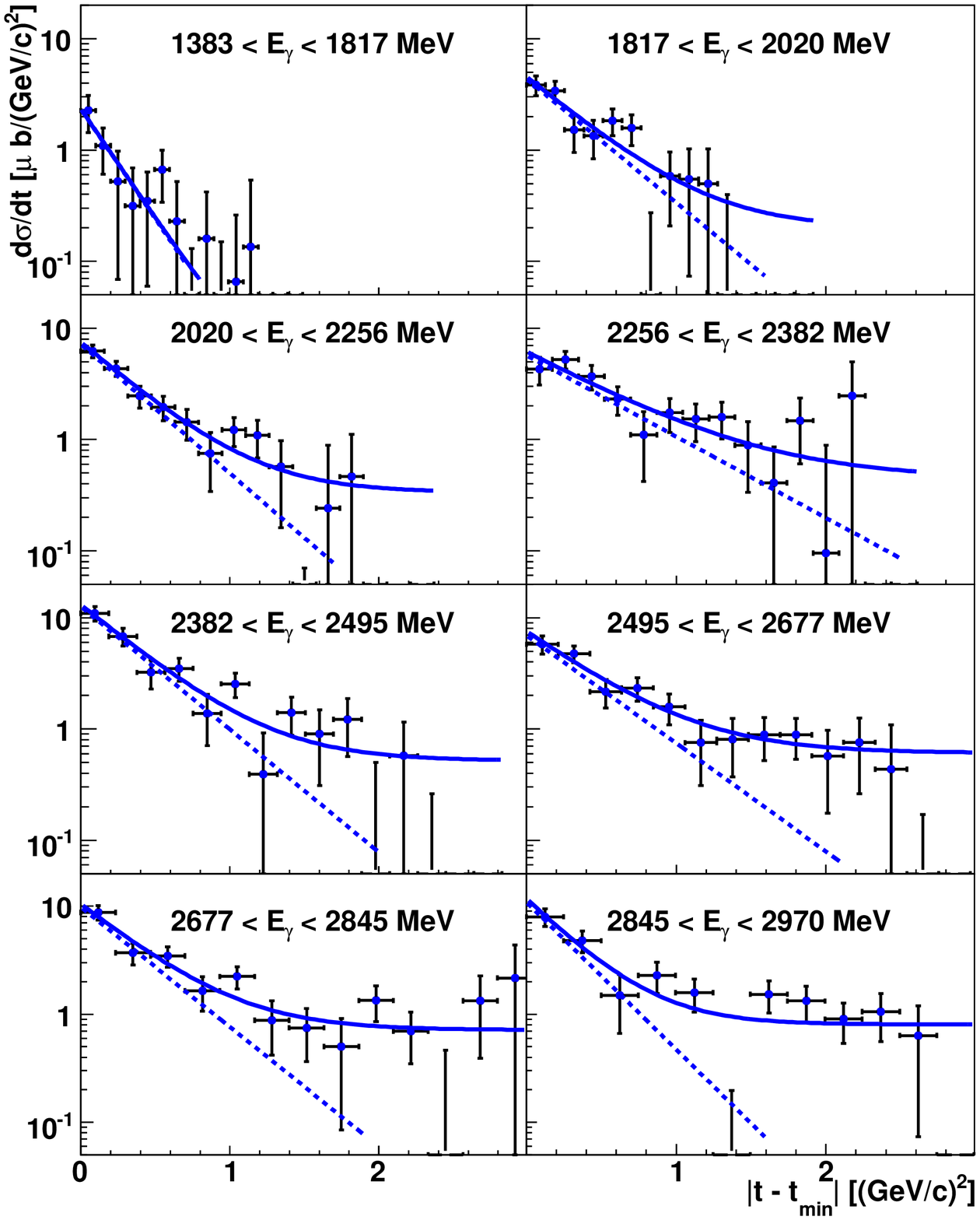}
\caption{Differential cross-sections $\dsdttext (| t - t_{\min}|)$ of
the squared four-momentum transfer $t$ to the $\proton\piz$ system. }
\label{fig:cs.tomega}
\end{figure}

Fig.~\ref{fig:cs.tomega} shows the differential cross-sections
$\dsdttext$ in dependence of $|t - t_{\min}|$, %for the $\omega$ meson,  
which are compatible with an exponential
behaviour in the low $t$ region. This is characteristic for production via $t$-channel
exchange. The data were fitted in the region below $0.8\cdot |t_{\max}
- t_{\min}|$ (approximately corresponding to $\epsilon > 5\,\%$) using
\begin{equation} \dsdt (|t - t_{\text{min}}|) =
e^{a + b|t - t_{\text{min}}| } + c(E)\,
\label{eqn:fit.t}
\end{equation}
where $t_{\text{min}}$ is the minimum squared momentum transfer imposed
by kinematics. The non-$t$-dependent contribution was described with
a function $c(E) = c_0 + c_1 \cdot E_{\gamma}$. The parameters $c_0$ and
$c_1$ were determined in a combined fit of the differential cross
sections. 

\begin{figure}[t]
\vspace{-4mm}
\begin{center}
\includegraphics[width=0.45\textwidth]{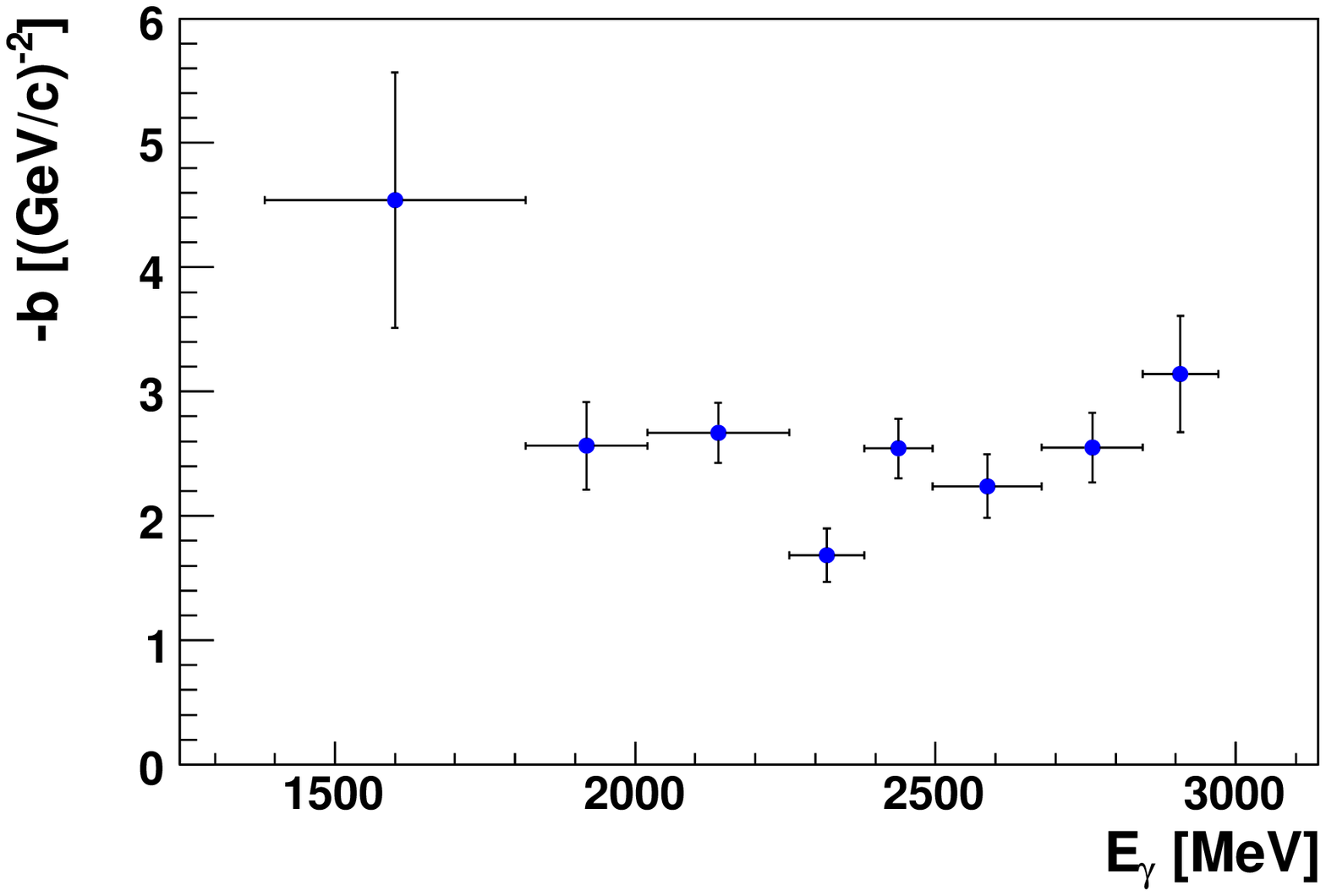} \end{center}
\caption{\label{fig:slope.omega}
Slope parameter of $\dsdttext (| t - t_{\min}|)$ }
\end{figure}

 The slope parameter $b$ is shown in Fig.~\ref{fig:slope.omega}. The
slope is approximately constant over the covered energy range. 
This indicates a strong contribution from 
$\omega$  production via $t$-channel exchange processes. 

From these differential distributions 
the total cross-section was obtained
by integrating over function~(\ref{eqn:fit.t}) from $t_{\min}$ to
$t_{\max}$.

\subsection{Total cross-section}
\label{sec:cs.total}

\begin{figure}[b]
\begin{tabular}{cc}
\hspace{-4mm}\includegraphics[width=0.25\textwidth]{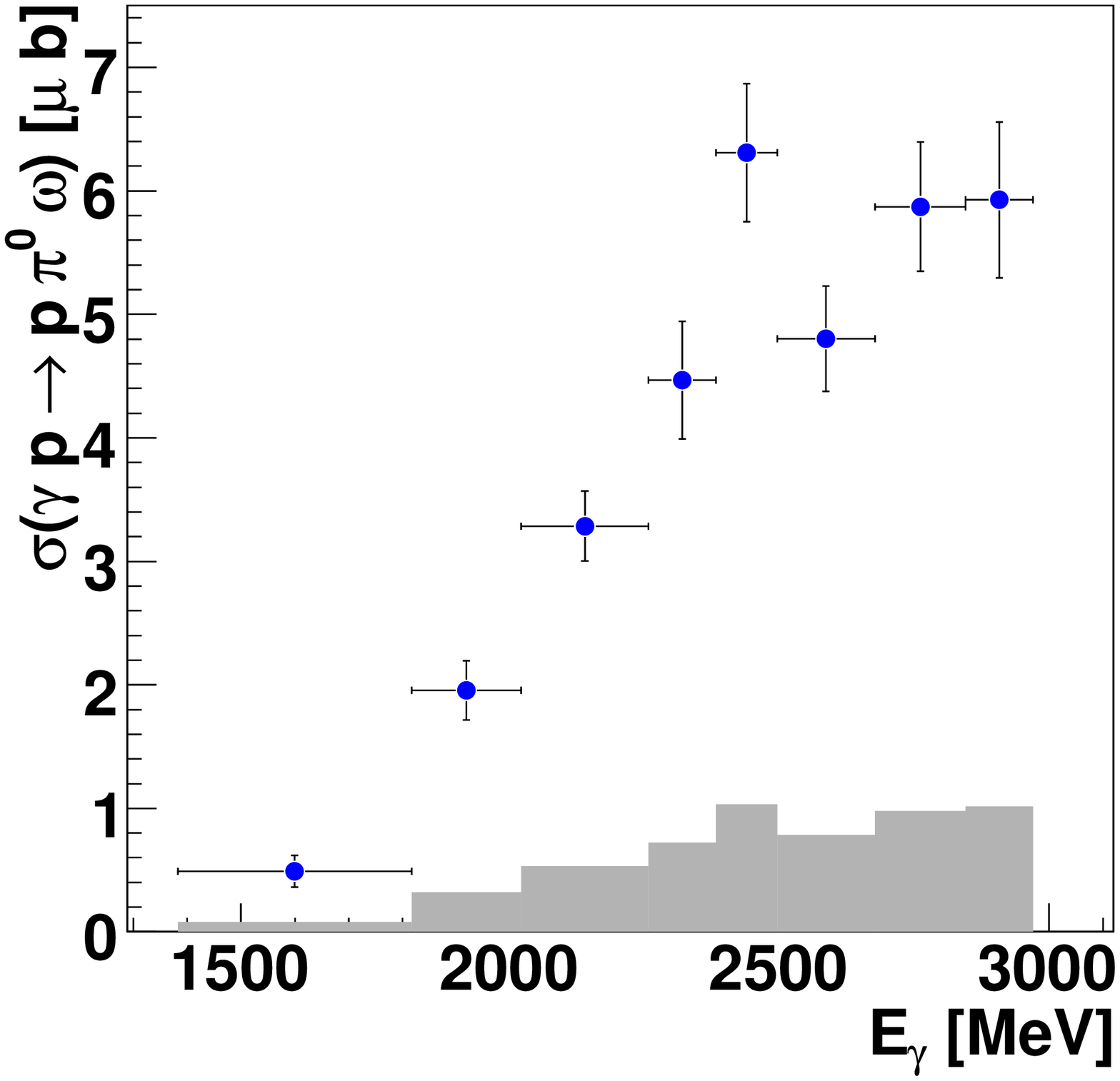} &
\hspace{-4mm}\includegraphics[width=0.25\textwidth]{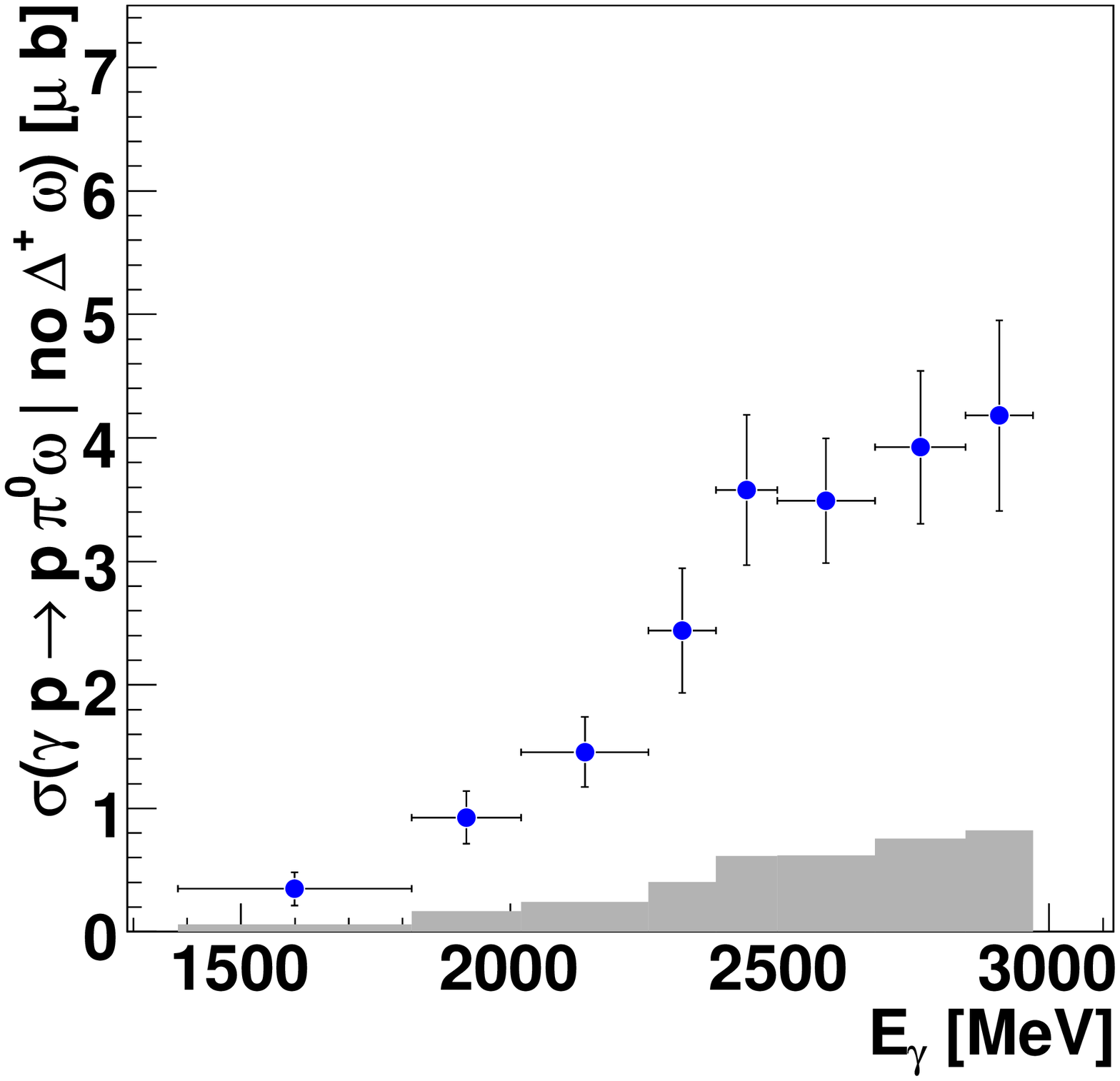}
\end{tabular}
\caption{\label{fig:cs.ppizomega}
Total cross-sections $\sigma(\gpppizomega)$ before (left) and after
(right) subtraction of the $\rm\Delta^{+}\omega$ contribution. The grey
band represents the systematic uncertainty. 
} \end{figure}

The total cross-section was determined in three different ways, by
extrapolation and summation of the three types of differential cross
sections, $\dsdomegatext (\cos\theta_{\omega})$, $\dsdomegatext (\cos
\theta_{\piz})$, and $\dsdttext (| t - t_{\min}|)$ as described above.
Statistical errors of the total cross-sections were determined by error
propagation. As final result, the mean value of the total cross-section
and the mean statistical error are shown in Fig.~\ref{fig:cs.ppizomega}
(left) as a function of the photon energy.  
The cross-section rises with increasing photon energy, i.\,e. with the
available phase space. 

A  systematic uncertainty
was derived from the spread of the three different determinations of the total cross-section, using data of Fig.~\ref{fig:cs.cosThOmega},
\ref{fig:cs.cosThPi0} and~\ref{fig:cs.tomega}. 
A further error of $5.7\,\%$  was  assigned to the Monte Carlo reconstruction 
efficiency~\cite{Amsler:1993kg}. 
These contributions and  the  $15\,\%$ normalisation  error \cite{Pee:06} were added in quadrature to yield the total systematic error shown in 
Fig.~\ref{fig:cs.ppizomega}.

\subsection{The \boldmath$\rm\Delta^{+}\omega$\unboldmath\ contribution to
\boldmath$\rm p\pi^{0}\omega$\unboldmath }
\label{sec:deltaomega}
\begin{figure}[pt]
\includegraphics[width=0.5\textwidth]{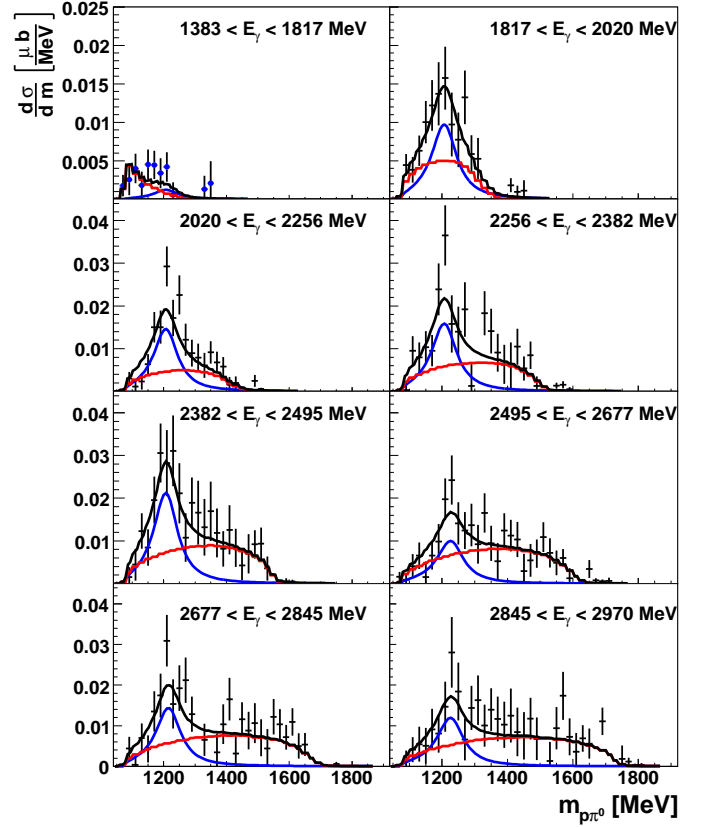}
\caption{Differential cross-sections $\dsdmtext
(\proton\piz)$. They are fitted with a combination of a Breit-Wigner (blue) and phase space $\ppizomega$ Monte Carlo events (red). } \label{fig:cs.mppi0}
\end{figure}

Fig.~\ref{fig:cs.mppi0} shows the differential cross-sections
$\dsdmtext (\proton\piz)$, which were used to disentangle the
$\rm\Delta^{+}\omega$ contribution to the total cross-section. The
distributions show prominent $\rm\Delta$ signals. The $\rm\Delta$ peak
was fitted by a phase space corrected Breit-Wigner function (see
e.\,g.~\cite{Anisovich:2005} for details).  The non-resonant
$\ppizomega$ part was described by phase-space distributed
$\ppizomega$ Monte Carlo events. Only the amplitudes of the two
contributions were left free in the fit. The Breit-Wigner width of the
$\rm\Delta$ was fixed to $120\,\MeV/{\rm c}^2$  the mass was fixed to
$1232\,\MeV/{\rm c}^2$ for energies below $2500\,\MeV$ 
and set to
values between $1240$ and $1250\,\MeV/\text{c}^2$ for higher energies to improve the fit. 
With these
two components a good description of the $\rm\Delta$ peak and of
the $\proton\piz\omega$ phase space contribution to the differential cross-section
was achieved.

\begin{figure}[pt]
\begin{center}
\includegraphics[width=0.47\textwidth]{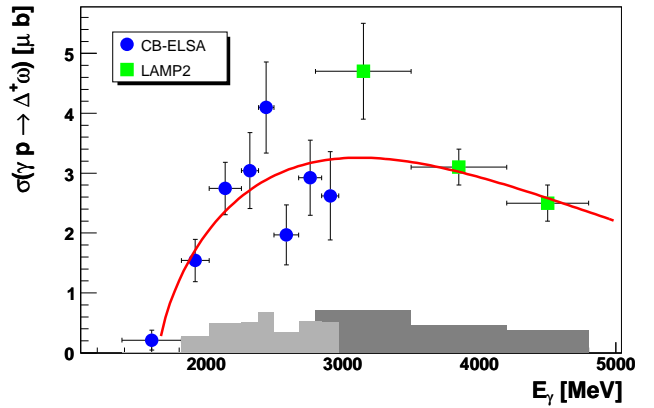}
\end{center}
\caption{\label{fig:cs.deltaomega}
Total cross-section $\sigma(\gpdeltaomega)$, shown are the data from this analysis and
from the LAMP2 experiment~\cite{Barber:1984fr}. The systematic errors are shown as an error band in light (CB-ELSA) and dark grey (LAMP2). 
A fit to the data points is shown, which is described in the text. 
} \end{figure}

The Breit-Wigner distributions and the $\proton\piz\omega$ phase-space
contributions were integrated and their fractions determined. The systematic uncertainty due
to the disentanglement was estimated to $3 - 10\,\%$ and added in quadrature to the systematic error.
The
cross-section for $\gpppizomega$ without $\rm\Delta^+\omega$
contributions is shown in Fig.~\ref{fig:cs.ppizomega} (right).

The total cross-section of $\gpdeltaomega$ was determined from the observed 
fraction of $\rm\Delta^+\omega$ events and the $\gpppizomega$ cross-section, 
taking into account the unseen  $\rm\Delta^+\to\neutron\pi^+$ decay mode. 
The
resulting cross-section is shown in Fig.~\ref{fig:cs.deltaomega}
together with the results of the LAMP2 experiment~\cite{Barber:1984fr}.
It is worthwhile to discuss how the LAMP2 cross-section was determined.

\begin{figure}[tb]
\begin{center}
\includegraphics[width=0.4\textwidth,height=45mm]{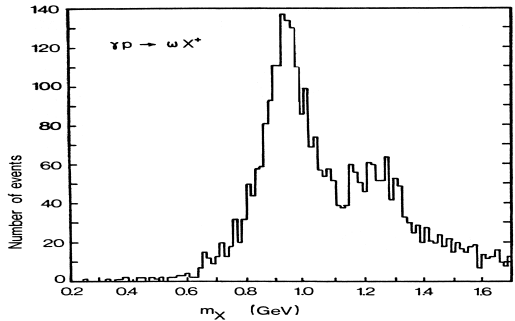}
\end{center}
\caption{\label{fig:MM.Lamp} Missing mass of the $\omega$ in the
LAMP2 experiment (from \protect\cite{Barber:1984fr}). Note that the
spectrum is dominated by $\proton\omega$ events.}
\end{figure}

The LAMP2 experiment measured the reaction $\gpdeltaomega$ by
identifying $\omega$ mesons in their $\pi^{+}\pi^{-}\piz$ decays. The
$\rm\Delta^{+}$ decay products were not observed. Instead, the
$\rm\Delta^{+}$  was identified in the missing mass spectrum of the
$\rm\gamma p\to\omega X$ reaction. The missing mass distribution
(Fig.~\ref{fig:MM.Lamp}) contains  signals for $\proton\omega$ and
$\rm\Delta^{+}\omega$ production. The authors give a $15\,\%$
systematic uncertainty due to the difficulty in disentangling the
$\proton\omega$, $\ppizomega$ and $\rm\Delta^{+}\omega$ contributions.

In our analysis the fraction of $\proton\piz$ below the $\rm\Delta$ is significant
(see fig.~\ref{fig:cs.mppi0}) and larger than estimated by LAMP2.
Hence it seems possible that the LAMP2 cross-section is overestimated.

The total cross-section for $\rm\Delta^{+}\omega$ photoproduction in Fig.
\ref{fig:cs.deltaomega} (see table~\ref{tab:cs.total}) is consistent with a simple fit assuming
a background amplitude in the form $A \cdot (E-E_{\rm threshold})^{\alpha/2}
\cdot (E-E_{\rm h})^{\beta/2}$ ($A$, $\alpha$, $\beta$, $E_{\text{threshold}}$, $E_{\rm h}$ fit
parameters). The $\chi^2=12.1$ for $N_{\text{DoF}} =  6$ corresponds
to an acceptable $6\,\%$ probability. 

\begin{table}[t]
\caption{\label{tab:cs.total} Total cross-sections of $\gpppizomega$ (with and without $\rm\Delta^+\omega$ contribution) and $\gpdeltaomega$.
 The systematic error is shown in Fig.~\ref{fig:cs.ppizomega} and~\ref{fig:cs.deltaomega}.}
\begin{center}
\begin{tabular}{cccc}
\hline
$E_{\gamma}$ & \hspace{-2mm}$\sigma(\gpppizomega)$\hspace{-2mm} & \hspace{-2mm}$\sigma(\gpppizomega)$\hspace{-2mm}& \hspace{-2mm}$\sigma(\gpdeltaomega)$\hspace{-2mm}\\
$[\MeV]$     & $[\ub]$ & (no $\rm\Delta^+\omega$)  $[\ub]$& $[\ub]$\\
\hline
$ 1383 - 1817$ &  $ 0.49 \pm 0.13$ & $ 0.35 \pm 0.13$ & $ 0.21 \pm 0.17$ \\ 
$ 1817 - 2020$ &  $ 1.95 \pm 0.24$ & $ 0.93 \pm 0.21$ & $ 1.54 \pm 0.35$ \\ 
$ 2020 - 2256$ &  $ 3.28 \pm 0.28$ & $ 1.46 \pm 0.28$ & $ 2.74 \pm 0.44$ \\ 
$ 2256 - 2382$ &  $ 4.47 \pm 0.47$ & $ 2.44 \pm 0.50$ & $ 3.04 \pm 0.63$ \\ 
$ 2382 - 2495$ &  $ 6.31 \pm 0.56$ & $ 3.58 \pm 0.61$ & $ 4.10 \pm 0.76$ \\ 
$ 2495 - 2677$ &  $ 4.80 \pm 0.43$ & $ 3.49 \pm 0.50$ & $ 1.97 \pm 0.50$ \\ 
$ 2677 - 2845$ &  $ 5.87 \pm 0.52$ & $ 3.92 \pm 0.62$ & $ 2.92 \pm 0.63$ \\ 
$ 2845 - 2970$ &  $ 5.93 \pm 0.63$ & $ 4.18 \pm 0.77$ & $ 2.62 \pm 0.74$ \\
\hline
\end{tabular}
\end{center}
\end{table}

\section{Summary}
\label{sec:summary}
We have studied the reaction $\gpppizomega$ with $\omega\to\pi^0\gamma$
from the $\omega\pi^0$ production threshold up to 3\,GeV
photon energy using an unpolarised tagged photon beam and a liquid
hydrogen target. Differential cross-sections were determined as
functions of $\cos\theta_{\omega}, \cos\theta_{\piz}$ and $| t -
t_{\min}|$. The distributions reveal strong contributions from isovector
exchange currents from the photon -- converting to an $\omega$ meson --
to the proton which undergoes a $\proton$-$\rm\Delta$ excitation. The cross
section for $\rm\Delta\omega$ production and for non-$\rm\Delta\omega$
events rises with increasing phase space; LAMP2 data indicate a
decrease of the cross-sections when going to larger photon energies
(3 - 5\,GeV). 

\begin{acknowledgement}
We thank the technical staff at ELSA and at all the participating
institutions for their invaluable contributions to the success of the
experiment. We acknowledge financial support from the Deutsche
Forschungsgemeinschaft (DFG) within the Sonderforschungsbereich
SFB/TR16. The collaboration with St. Petersburg received funds from DFG
and the Russian Foundation for Basic Research. \mbox{B.~Krusche}
acknowledges support from Schweizerischer Nationalfond. U.~Thoma thanks
for an Emmy Noether grant from the DFG. A.\,V.~Anisovich and
A.\,V.~Sarantsev acknowledge support from the Alexander von Humboldt
Foundation. This work comprises part of the PhD thesis of J.\
Junkersfeld.  %and J.\ Lotz.
\end{acknowledgement}

% BibTeX users please use

%\bibliographystyle{unsrt}
%\bibliography{lit_jj}

\end{document}